\documentclass[12pt]{amsart}
\usepackage{amsthm,amsmath,amsxtra,amscd,amssymb,xypic,color,xspace,esint}
\usepackage{tikz}
\usepackage{latexsym,amsmath,amssymb,mathrsfs}
\usepackage{amsfonts,eucal,amsthm,graphicx,color}

\numberwithin{equation}{section}

\newcommand{\ve}{{\varepsilon}}

\def\beq{\begin{equation}}
\def\eeq{\end{equation}}

\newcommand\tr{{\rm tr}}

\theoremstyle{plain}

\theoremstyle{definition}

\newenvironment{aside}{\begin{quote}\sffamily}{\end{quote}}

\newcommand {\BC}   {\mathbb C}

\newcommand {\BR}   {\mathbb R}

\newcommand {\qe} {\mathfrak q}

\newcommand {\BS}   {\mathbb S}

\newcommand {\BZ}   {\mathbb Z}

\newcommand {\CalH} {\mathcal H}

\newcommand {\CalN} {\mathcal N}

\newcommand {\CalT} {\mathcal T}

\newcommand {\CalV} {\mathcal V}

\newcommand{\fo}{\vert\kern -.03in\_}

\newcommand {\ii} {\mathrm{i}}
\newcommand{\Tr}{\text{Tr}}
\newcommand{\pa}{{\partial}}

\begin{document}
\title{Superspin chains\\  
$\uparrow$ {\tt and} $\downarrow$\\
 supersymmetric gauge theories}

\author{Nikita Nekrasov}
\address{Simons Center for Geometry and
    Physics, Stony Brook University, Stony Brook, NY 11794\newline
    Kharkevich Institute for Information Transmission Problems, Moscow 127051 Russia\newline Center for Advanced Studies,
Skolkovo Institute of Science and Technology,
1 Nobel Street, Moscow,  143026 Russia  
\newline
    e-mail: nikitastring@gmail.com
    }
\date{}

\maketitle

\emph{To Martin Rocek on his super-anniversary, with love}

\begin{abstract}
 
 We discuss the possible extensions of Bethe/gauge correspondence to quantum integrable systems based on the super-Lie algebras of $A$ type. Along the way we propose the analogues of Nakajima quiver varieties whose cohomology and K-theory should carry the representations of the corresponding Yangian and the quantum affine algebras, respectively. We end up with comments on the ${\CalN}=4$ planar super-Yang-Mills theory in four dimensions.  

\end{abstract}

\section{Introduction}

Gauge theories with ${\CalN}=(2,2)$ super-Poincare symmetry have an interesting connection to quantum integrable systems. Perhaps the first instance of such a connection has been spotted in the studies of the two dimensional Yang-Mills theory \cite{Migdal}, interpreted \cite{Witten:2d}
as a topological field theory, which can by obtained \cite{Witten:1992xu} from a twisted version of the ${\CalN}=(2,2)$ theory by a (non-unitary) deformation, namely one turns on the twisted superpotentials ${\tilde W}$ and ${\tilde W}^{*}$ which are not complex conjugate. The expression \cite{Witten:1992xu} for the partition function of the theory on a compact Riemann surface makes it clear the physical states  of the topological  theory (which are the vacua of the supersymmetric theory) are in one-to-one correspondence with the states of a free particle living on the space of conjugacy classes $T/W$ of the gauge group $G$. For $G = SU(N)$ this system is equivalent to the system of free $N$ fermions on living on a circle. In \cite{Gorsky:1993pe} this relation has been generalized to allow for certain line operators in gauge theory. In the presence of line operators the formerly free fermions become interacting, but they dynamics remains integrable. The energy eigenvalues of the many-body system is identified with the vacuum expectation value of the local observable ${\Tr} {\sigma}^{2}$, where $\sigma$ is the complex adjoint scalar in vector multiplet. In \cite{MNS} the example of \cite{Gorsky:1993pe} has been upgraded: one studied the two dimensional (twisted) ${\CalN}=(2,2)$ $SU(N)$ gauge theory with adjoint chiral multiplet, of twisted mass \cite{AlvarezGaume:1983ab}  $c$, and discovered that the vacua were in one-to-one correspondence with the stationary states of a system of $N$ particles $x_{1}, \ldots , x_{N}$ on a circle, interacting with the repulsive potential $c {\delta}(x_{i}-x_{j})$. This example has been further explored in \cite{GS}. Then, in \cite{Nekrasov:2009uh} the general correspondence has been identified: supersymmetric vacua of gauge theories with ${\CalN}=(2,2)$ $d=2$ Poincare supersymmetry (the theories need not be two dimensional) are the stationary states of some quantum integrable system, i.e. they are the joint eigenvectors of quantum integrals of motion. Moreover, this correspondence  has a remarkable social feature: the textbook examples of supersymmetric gauge theories map to the textbook examples of quantum integrable systems. A large class of models has been found where the quantum integrable system is based on quantum algebras of the $A,D,E$-type, such as the spin chains with the corresponding spin group. The dual gauge theory is of the $A,D,E$ quiver-type. The mathematical consequence of this relation is the connection \cite{Nekrasov:2009uh} between quantum groups: Yangians, quantum affine algebras, elliptic quantum groups,  and quantum cohomology, quantum K-theory, and elliptic cohomology, respectively. In the series of remarkable works  \cite{MO, Okounkov:2015spn,Aganagic:2016jmx} this connection has been elucidated and put on the firm mathematical ground, moreover, for general quivers, not only of the (affine) $A,D,E$ type. On the physics side the quiver gauge theories in question are softly broken ${\CalN}=(4,4)$ theories (in two dimensions). The parameter of deformation, the twisted mass corresponding to a specific $U(1)$ R-symmetry, maps to the Planck constant of an integrable system.  

In this paper we attempt to extend the realm of the correspondence to the case of super-algebra based quantum integrable systems. We should point out that gauge theories based on supergroups naively make no sense, as the invariant scalar product on the Lie superalgebra is not positive definite, so the theory is not unitary. Nevertheless, the supergroup gauge symmetry is possible in the context of topological field theory, such as Chern-Simons theory in three dimensions, albeit there are caveats \cite{Vafa:2001qf, Mikhaylov:2015nsa, Mikhaylov:2014aoa}. Also, the analytic continuation of a conventional gauge theory may reach the supergroup gauge theory \cite{Dijkgraaf:2016lym}. 

Our motivation also includes the desire to get a better understanding of the integrable structure behind the planar limit of ${\CalN}=4$ super-Yang-Mills theory in four dimensions. It has been discovered, first in a $SU(2)$ sector \cite{Minahan:2002ve} and then in the general case
  \cite{Beisert:2003yb, Beisert:2004hm, Beisert:2006ez, Beisert:2006ib}, that the spectrum of anomalous dimensions of local operators is that of a quantum integrable spin chain based on the Yangian $Y(\mathfrak{gl}(4|4))$of the superconformal group 
, see the excellent review in \cite{Drummond:2010km}. For most of the integrable spin chains the Bethe equations can be cast in the form:
\beq
{\exp}\, \frac{\partial {\tilde W}}{\partial \sigma_{i}} = 1 \, ,  \qquad i = 1, \ldots, M
\label{eq:bap}
\eeq
 where  $\sigma_i$ are the Bethe roots, and ${\tilde W}$ is the so-called Yang-Yang function. It can be shown, however, that the dressing phase
 \cite{Beisert:2006ib} entering the Bethe equations in the ${\CalN}=4$ super-Yang-Mills and \cite{Arutyunov:2004vx} on the $AdS_{5} \times {\BS}^{5}$ dual side, violates the potentiality of \eqref{eq:bap}. 
 Despite many works explaining the origin of the dressing phase and investigating its analytic structure, e.g.\cite{Sakai:2007rk,Gromov:2007fn, Dorey:2007xn} the satisfying explanation on the side of the supersymmetric gauge theory with ${\CalN}=(2,2)$ supersymmetry in two dimensions is missing. The explanation might be the further breaking of supersymmetry $(2,2) \to (0,2)$ \cite{NS}. 

In this paper we make a modest step in this direction. We shall propose a class of ${\CalN}=(2,2)$ quiver gauge theories in two dimensions, whose supersymmetric vacua are in one-to-one correspondence with the Bethe states of closed spin chains based on the Yangian of $\mathfrak{gl}(m|n)$.

The paper is organized as follows. The section $\bf 2$ starts with review of the simplest example of Bethe/gauge correspondence, where the quantum integrable system is the Heisenberg spin chain, while the supersymmetric gauge theory is the gauged linear sigma model with the target space being the cotangent bundle to the Grassmanian of $N$-dimensional planes $V$ in the $L$-dimensional complex vector space $W$. We recall Bethe equations, their Yang-Yang form, and the $T-Q$ equation which is equivalent to them. We also briefly review the generalizations: to other spin groups, to inhomogeneous, twisted and anisotropic cases. The section $\bf 3$ reviews Bethe equations for the superspin chains, based on $\mathfrak{gl}(M|N)$ algebra. The section $\bf 4$ introduces the main character: the gauge theory with the proper structure of its supersymmetric vacua. We'll see that Bethe equations themselves do not fix the matter content uniquely. We shall propose a  family of theories, $L_{\vec t}$ with the parameters $\vec t$ being the mass terms in the superpotential. The ${\vec t} =0, \infty$ theories can be topologically twisted so as to define an $A$-model. The intermediate theories flow, in the infrared, to the ${\vec t} = \infty$ point. However, we believe it is the ${\vec t}=0$ which should be identified with the Bethe/gauge dual of the superspin chain, as the ${\vec t} = \infty$ being effectively a theory with fewer fields, is less rigid, and, in fact, has additional marginal deformation, which masks the Planck constant. The section $\bf 5$ concludes with unfinished business and future directions. 
 
\subsubsection{Acknowledgement} Research was partly supported by RFBR grant 18-02-01081.  
I am grateful to M.~Aganagic, V.~Kazakov, A.~Okounkov, V.~Pestun, S.~Sethi, A.~Tseytlin and especially E.~Ragoucy for discussions. Part of this work was done while I visited Physics Department at Imperial College London in 2013-2017, and IHES in 2013-2018.  The results were presented at the 2018 MSRI program ``Enumerative geometry beyond numbers'' organized by M.~Aganagic and A.~Okounkov and the 2018 Royal Society meeting ``Quantum integrability and quantum Schubert calculus'' organized by V.~Gorbounof and C.~Korff. I thank  them  for their hospitality. 

Finally, I would like to thank Martin Rocek for all his help and for all the conversations we had (and hopefully will have) on life and physics, supersymmetric or not so much.

\section{Heisenberg, Bethe, and Grassmann}

\subsection{Spin chain}

The Heisenberg spin chain
\beq
{\hat H} = \sum_{a=1}^{L} {\vec\sigma}_{a} \otimes {\vec\sigma}_{a+1} \ ,
\label{eq:heisham}
\eeq
where 
\beq
{\vec\sigma}_{a+L} = {\vec\sigma}_{a}
\ ,
\label{eq:per}
\eeq 
has an $SU(2)$ underlying symmetry: $\vec\sigma_{a} = \left( {\sigma}^x_a, {\sigma}^y_a, {\sigma}^z_a \right)$ are the generators of $SU(2)$ acting at the site $a$ of the length $L$ spin chain.  The eigenvectors 
\beq
{\psi} \in \left( {\BC}^{2} \right)^{\otimes\, L} = \bigoplus_{N=0}^{L} {\CalH}_{N}\, , \eeq 
\beq
{\rm dim}_{\BC} {\CalH}_{N} = \left( \begin{matrix} L \\ N \end{matrix} \right)
\eeq are constructed, in the Bethe ansatz approach, from the solutions of the Bethe ansatz equations:
\beq
\prod_{\alpha \neq \beta} \frac{{\sigma}_{\beta} - {\sigma}_{\alpha} + 2u}{{\sigma}_{\beta} - {\sigma}_{\alpha} - 2u} = 
\left( \frac{{\sigma}_{\beta} +u}{{\sigma}_{\beta} - u} \right)^{L} \, , \qquad {\beta} = 1, \ldots , N
\label{eq:bae1}
\eeq
which can be, equivalently, represented via the so-called T-Q equation:
\beq
P(x-u) Q(x+2u) + P(x+u) Q(x-2u) = T(x) Q(x)
\label{eq:tq1}
\eeq
where
\beq
Q(x) = \prod_{{\beta}=1}^{N} ( x - {\sigma}_{\beta} ) \, , \qquad 
\label{eq:qop}
\eeq
\beq
P(x) = x^{L} \, , \
\label{eq:ad1}
\eeq
and $T(x)$ is some polynomial of degree $L$. Finally, with the help of the $Y$-observable:
\beq
Y(x) = \frac{Q(x)}{Q(x-2u)}
\eeq
one rewrites \eqref{eq:tq1} as:
\beq
Y(x+2u) + D(x) \frac{1}{Y(x)} = \frac{T(x)}{P(x-u)}\, , \qquad D(x) = \frac{P(x+u)}{P(x-u)}
\label{eq:qchar1}
\eeq
the content of this equation being the absence of the poles of the left hand side in $x$, other then zeroes of $P(x-u)$. 
 All this generalizes in a relatively straightforward way, both in terms of the spin group symmetry, and the possibilities of the choice of the Hamiltonian. Recall three upgrades:
 twisting, inhomogeneity and anisotropy. The first two don't change the underlying symmetry generating algebra, while the last one deforms the rational algebra (the Yangian) into the quantum affine and elliptic quantum algebras, respectively. 
 
The inhomogeneity deforms the Hamiltonian \eqref{eq:heisham} in certain fashion, making the spin interactions, in general, $a$-dependent,  and less local, 
while twisting deforms the boundary conditions \eqref{eq:per} to
\beq
{\vec \sigma}_{a+L} = {\qe}^{-\frac{\ii \sigma_3}{2}} {\vec \sigma}_{a}  {\qe}^{\frac{\ii \sigma_3}{2}}
\label{eq:twper}
\eeq
Both deformations preserve integrability. The only aspect of these deformations needed for the Bethe/gauge correspondence is their impact on Bethe equations: the Eqs. \eqref{eq:bae1} deform to
\beq
\prod_{{\beta}' \neq {\beta}} \frac{{\sigma}_{\beta} - {\sigma}_{\beta'} + 2u}{{\sigma}_{\beta} - {\sigma}_{\beta'} - 2u} \ = \ {\qe}
\prod_{a=1}^{L} \frac{{\sigma}_{\beta} +u- {\mu}_{a}}{{\sigma}_{\beta} - u- {\mu}_{a}}  \, , \qquad {\beta} = 1, \ldots , N
\label{eq:bae2}
\eeq
where
\beq
P(x) = \prod_{a=1}^{L} (x - {\mu}_{a})\, , 
\label{eq:pp2}
\eeq
while \eqref{eq:tq1} deforms to 
\beq
P(x-u) Q(x+2u) + {\qe} P(x+u) Q(x-2u) = (1+{\qe}) T(x) Q(x)
\label{eq:tq2}
\eeq
and \eqref{eq:qchar1} to
\beq
Y(x+2u) + {\qe} D(x) Y(x)^{-1} = (1+{\qe}) T(x)/P(x-u)
\eeq 

\subsection{Gauge theory}

The gauge theory for which \eqref{eq:bae2} describe its vacua, is the softly broken ${\CalN}=(4,4)$ supersymmetric gauge theory in two dimensions, with the gauge group $U(N)$, and $L$ hypermultiplets in fundamental representation. Viewed as an ${\CalN}=(2,2)$ theory, it has a vector multiplet $(A_{m}, {\sigma})$, an adjoint-valued chiral multiplet ${\Phi}$, and $L$ pairs of chiral multiplets $(Q_{a}, {\tilde Q}^{a})$, $a = 1, \ldots , L$, with $Q_{a} = (Q_{a}^{\beta})_{{\beta}=1}^{N}$ transforming in the fundamental $N$-dimensional representation ${\bf N}$ of $U(N)$, ${\tilde Q}^{a} =({\tilde Q}^{a}_{\beta})_{{\beta}=1}^{N}$ transforming in the conjugate representation $\bar{\bf N}$. In addition, the theory has a superpotential $W = \sum_{a=1}^{L} {\tilde Q}^{a}{\Phi}Q_{a}$, and the twisted masses $u, {\mu}_{a}$, corresponding to the $U(1)_{u} \times U(L)$ global symmetry: 
$U(L)$ acts on ${\tilde Q}$ in the $L$-dimensional fundamental representation $\bf L$, on $Q$ in the conjugate $\bar{\bf L}$. The 
$U(1)_{u}$ symmetry acts via: $({\Phi}, Q, {\tilde Q}) \mapsto ({\Phi}e^{2\ii \alpha} , Q e^{-\ii\alpha} , {\tilde Q} e^{-\ii\alpha})$. 
The list of relevant parameters of the theory concludes with the Fayet-Illiopoulos parameter $r$ and the abelian $\theta$-angle, which are conveniently combined into
\beq
{\qe} = e^{2\pi \ii \theta} e^{-r}
\label{eq:kahler1}
\eeq
Suppose we are in the phase where the complex adjoint scalar $\sigma$ in the vector multiplet has the vacuum expectation value ${\sigma} = {\rm diag} ({\sigma}_1 , \ldots , {\sigma}_{N})$, as dictated by the potential ${\tr} \left( [{\sigma}, {\sigma}^{\dagger}] \right)^{2}$. The physical masses of the matter fields are: 
\beq
\begin{aligned}
& |{\sigma}_{\beta} - {\sigma}_{\beta'} +2u| \, , \qquad {\rm  for}  \ {\Phi}_{\beta}^{\beta'} , 
\\
& |{\sigma}_{\beta} - {\mu}_{a} -u |\, , \qquad {\rm for}  \ Q_{a}^{\beta} , \\
& |{\mu}_{a} - {\sigma}_{\beta}  - u|\, , \qquad {\rm  for} \ {\tilde Q}^{a}_{\beta}  \ . \\
\label{eq:physma}
\end{aligned}
\eeq
Assuming they are all non-zero we integrate out the matter fields and the non-abelian degrees of freedom in the vector multiplet (these have masses $\sim |{\sigma}_{\beta}  - {\sigma}_{\beta'}|$) to produce the effective twisted superpotential 
\beq
{\tilde W} = {\tilde W}^{\rm tree} + {\tilde W}^{\rm 1-loop} \, , 
\eeq
where
\beq
{\tilde W}^{\rm tree} = \frac{{\rm log}({\qe})}{2\pi\ii} \sum_{i=1}^{N} {\sigma}_{i}
\eeq
and, with ${\varpi}(x) = \frac{x}{2\pi \ii} ( {\rm log}(x) - 1 )$, 
\begin{multline}
{\tilde W}^{\rm 1-loop} = \sum_{\rm fields} {\varpi} (Q_{\rm field} )  = \\
\sum_{\beta,\beta'}  {\varpi} ({\sigma}_{\beta} - {\sigma}_{\beta'} + 2u) + 
\sum_{\beta,a} \left( {\varpi} ({\sigma}_{\beta} - {\mu}_{a} - u) + {\varpi} ( {\mu}_{a} - {\sigma}_{\beta} - u ) \right) 
\label{eq:1loop}
\end{multline}
The specific feature of the twisted superpotential, as opposed to the more familiar superpotential, is the multivaluedness of its first derivative, which is related to the discrete nature of the top component $F_{i}$ of the twisted chiral superfield $\Sigma_{\beta}  ={\sigma}_{\beta} + \ldots + {\vartheta}{\tilde\vartheta} F_{\beta}$ which enters the Lagrangian of the effective theory through the twisted $F$-term $\int d{\vartheta}d{\tilde\vartheta} {\tilde W} ({\Sigma})$. The minima of the effective potential (which involves the coupling to the field strengths $(F_{\beta})_{i=1}^{N}$ of the abelian gauge fields) are the solutions to the equations:
\beq
{\exp} \, 2\pi \ii \frac{\partial {\tilde W}}{\partial \sigma_{\beta}} \ = \ 1 \, , \ {\beta} = 1, \ldots , N
\label{eq:susyvac}
\eeq
which happily coincide with \eqref{eq:bae2}. As long as the masses of the matter fields \eqref{eq:physma} as well as those of the $W$-bosons are non-zero, the exactness of the one-loop approximation \eqref{eq:1loop} can be justified. 

The implications of the identification of \eqref{eq:susyvac} with \eqref{eq:bae2} are quite dramatic. One of the unexpected consequences is the realization that the Yangian of $\mathfrak{sl}_{2}$, which is the spectrum generating algebra of the Heisenberg spin chain, must act in the union of Hilbert spaces of \emph{different} quantum field theories, namely $U(N)$ gauge theories with all values of $N$, at least for $N \leq L$.  The specific realization of this novel symmetry is not yet completely understood, although the constructions of \cite{Nakajima, Varagnolo,MO} provide the tantalizing hints.

\subsection{Generalizations}

Let us now briefly review the generalization of the above correspondence to the case of a Lie algebra $\mathfrak{g}_{Q}$ based on a quiver $Q$. The vertices $v \in V_{Q}$ are the simple roots while the edges connecting the vertices encode their scalar products. The simple Lie algebras $\mathfrak{sl}_{r+1}$, $\mathfrak{so}_{2r}$, $\mathfrak{e}_{r}$ with $r= 6, 7,8$ and their affine versions are associated with the quivers with $r$ ($r+1$) vertices, which coincide with their  Dynkin diagrams. 

The spin chain model based on $\mathfrak{g}_{Q}$ depends on the choice of the representation ${\CalH}_{\bf w}$ of the Yangian $Y(\mathfrak{g}_{Q})$, which, in turn, can be taken to be the tensor product of the so-called evaluation representations $R_{i}({\mu})$, where $i \in V_{Q}$ and ${\mu} \in {\BC}$:
\beq
{\CalH}_{\bf w} = \bigotimes\limits_{i \in V_{Q}}  \, \bigotimes\limits_{{\alpha}=1}^{w_{i}} R_{i}({\mu}_{\alpha}^{(i)}) \, ,
\label{eq:repspin}
\eeq
where ${\mu}_{\alpha}^{(i)} \in {\BC}$. The multiplicities ${\bf w} = (w_{i})_{i \in V_{Q}}$ are the analogues of $L$, and the evaluation points ${\mu}_{a}^{(i)}$ are  the analogues of the parameters ${\mu}_{1}, \ldots , {\mu}_{L}$. Now, the analogue of the spin projection $N$ is the collection ${\bf v} = (v_{i})_{i\in V_{Q}}$, where $v_{i} \in {\BZ}_{\geq 0}$. 

The Bethe ansatz equations in the case of general $Q$ are sometimes called the nested Bethe equations (in the case of the $A, D, E$ Dynkin diagrams they were written in \cite{Reshet,KR}). The unknowns are the Bethe roots
${\sigma}_{\beta}^{(i)}$, where $\beta = 1, \ldots , v_{i}$, $i \in V_{Q}$. 
These equations have the Yang-Yang potential:
\begin{multline}
{\tilde W}_{Q}  = \frac{1}{2\pi \ii} \sum_{i \in V_{Q}} \ {\rm log}{\qe}_{i}\, \sum_{\beta = 1}^{v_{i}} {\sigma}_{\beta}^{(i)} + \\
 \sum_{i \in V_{Q}} \sum_{\beta = 1}^{v_{i}} \scriptstyle{\left(  \sum_{{\beta}' = 1}^{v_{i}} {\varpi} ( {\sigma}_{\beta}^{(i)} - {\sigma}_{\beta'}^{(i)} + 2u ) +\sum\limits_{a = 1}^{w_{i}} \left( {\varpi} ( {\sigma}_{\beta}^{(i)} - {\mu}_{a}^{(i)} - u ) +  {\varpi} ( - {\sigma}_{\beta}^{(i)} + {\mu}_{a}^{(i)} - u ) \right) \right)} + \\
 \sum_{e \in E_{Q}} \sum_{{\alpha} = 1}^{v_{s(e)}} \sum_{{\beta}=1}^{v_{t(e)}} \scriptstyle{\left( {\varpi} ( {\sigma}_{\alpha}^{(s(e))} - {\sigma}_{\beta}^{(t(e))} - u + {\mu}_{e} ) + {\varpi} ( - {\sigma}_{\alpha}^{(s(e))} + {\sigma}_{\beta}^{(t(e))} - u - {\mu}_{e} ) \right)}  \\
 \label{eq:genyy}
 \end{multline}
where, in order to write the equations, one introduces some orientation of the edges, thereby defining two maps $s,t: E_{Q} \to V_{Q}$, sending an edge $e \in E_{Q}$ to its source $s(e)$ and the target $t(e)$, respectively. 
The new entry in \eqref{eq:genyy} is a $\BC$-valued $1$-cochain $({\mu}_{e})_{e\in E_{Q}}$ which can be eliminated by redefining ${\mu}_{a}^{(i)}$'s for simply-connected $Q$'s. 
The observation of \cite{Nekrasov:2009uh} is that \eqref{eq:genyy} is precisely the effective twisted superpotential of the ${\CalN}=(4,4)$ 
theory in two dimensions with the gauge group
\beq
G_{\bf v} = \times_{i\in V_{Q}} \ U(v_{i})
\eeq
and the hypermultiplets in the representations 
\beq
R_{H} = \bigoplus\limits_{i \in V_{Q}}  Hom ({\bf w}_{i} , {\bf v}_{i}) \ 
\bigoplus\limits _{e \in E_{Q}} Hom ({\bf v}_{s(e)} , {\bf v}_{t(e)})
\eeq
where ${\bf w}_{i} \approx {\BC}^{w_{i}}$ are the multiplicity spaces, and
${\bf v}_{i} \approx {\BC}^{v_{i}}$ are the defining representations of $U(v_{i})$. The parameter $u$ is the twisted mass softly breaking the
supersymmetry down to ${\CalN}=(2,2)$, it corresponds to the $U(1)$ symmetry under which the ${\CalN}=2$ adjoint chiral multiplets $\Phi_{i}$ in ${\CalN}=4$ vector multiplets have charge $+2$, while the ${\CalN}=2$
chiral multiplets in fundamental $Hom ({\bf w}_{i} , {\bf v}_{i})$ and antifundamental $Hom ({\bf v}_{i} , {\bf w}_{i})$ representations, as well
as both bi-fundamentals $Hom ({\bf v}_{s(e)} , {\bf v}_{t(e)})$ and its conjugates $Hom ({\bf v}_{t(e)} , {\bf v}_{s(e)})$ have charge $-1$. The parameters ${\mu}_{e}$ are the twisted masses corresponding to the $U(1)_{e}$ symmetry under which $Hom ({\bf v}_{t(e)} , {\bf v}_{s(e)})$ has $+1$ charge, while $Hom ({\bf v}_{s(e)} , {\bf v}_{t(e)})$ has $-1$ charge. The evaluation parameters ${\mu}_{a}^{(i)}$ are the twisted masses for the maximal torus of $U(w_{i})$. 

\section{Bethe ansatz for closed super-spin chains}

The Bethe ansatz equations for the spin chains based on the superalgebra $\mathfrak{gl}(m|n)$ has been found long time ago. We use the formalism of \cite{Ragoucy:2007kg}  and \cite{Belliard:2008di}, adapted to our notations. 

\subsubsection{Principal gradation}

Let us first discuss the case of the principal gradation Dynkin diagram \cite{Frappat:1996pb}.

 The diagram has $n+m-1$ node, with  $i = 1, \ldots , m-1$ and $i=m+1, \ldots
 m+n-1$ called the bosonic nodes and $i=m$ the fermionic node. The Bethe roots ${\sigma}_{\alpha}^{(i)}$, ${\alpha} = 1, \ldots , v_{i}$
 are the roots of the polynomials $Q_{i}(x)$, $i = 1, \ldots , m+n-1$ of degrees $v_{i}$,
 \beq
 Q_{i}(x) = \prod\limits_{{\alpha}=1}^{v_{i}} \left( x - {\sigma}_{\alpha}^{(i)} \right) 
 \label{eq:qbaxi}
 \eeq
 We also define $Q_{0}(x) = Q_{m+n}(x) \equiv 1$. 
Then Bethe equations (we generalized them by including the twist parameters $\qe_i$'s) have the form: whenever $Q_{i}(x) = 0$:
\beq
\begin{aligned}
& \quad {\qe}_{i} \, \frac{Q_{i-1}(x + u)}{Q_{i-1}(x - u)}\frac{Q_{i}(x-  2u)}{Q_{i}(x + 2u)}\frac{Q_{i+1}(x + u)}{Q_{i+1}(x - u)} = - \frac{P_{i}(x + u)}{P_{i}(x - u)} \, ,  \qquad \scriptstyle{1 \leq i \leq m-1} \, , \\
& 
\qquad\qquad {\qe}_{m} \, \frac{Q_{m-1}(x + u)}{Q_{m-1}(x - u)}\frac{Q_{m+1}(x - u)}{Q_{m+1}(x + u)} = - \frac{P_{+}(x)}{P_{-}(x)}\, , \\
& \quad {\qe}_{i} \, \frac{Q_{i-1}(x - u)}{Q_{i-1}(x +  u)}\frac{Q_{i}(x +  2u)}{Q_{i}(x -  2u)}\frac{Q_{i+1}(x + u)}{Q_{i+1}(x - u)} = - \frac{P_{i}(x - u)}{P_{i}(x + u)} \, ,  \qquad \scriptstyle{m < i \leq m+n-1} \, , \\
\label{eq:baesusy}
\end{aligned}
\eeq
with monic polynomials $P_{k}(x)$, $k = 1, \ldots , m-1, \pm , m+1,  \ldots, m+n-1$. 
We see that \eqref{eq:baesusy} can be cast in the form 
\beq
{\exp}\, \frac{\pa {\tilde W}_{\mathfrak{gl}(m|n)}}{\pa \sigma_{\alpha}^{(i)}} = 1
\eeq
where ${\tilde W}_{\mathfrak{gl}(m|n)}$ is similar to the $Q = A_{m+n-1}$ Yang-Yang function \eqref{eq:genyy}, except that the node $i=m$ contributes differently, and the sign of $u$ is flipped in passing from $i < m$ to $i>m$:
\begin{multline}
{\tilde W}_{\mathfrak{gl}(m|n)}
 = \frac{1}{2\pi \ii} \sum_{i =1}^{m+n-1} \ {\rm log}{\qe}_{i}\, \sum_{\beta = 1}^{v_{i}} {\sigma}_{\beta}^{(i)} + \\
 \sum_{i =1}^{m-1} \sum_{\beta = 1}^{v_{i}} \scriptstyle{\left(  \sum\limits_{{\beta}' = 1}^{v_{i}} {\varpi} ( {\sigma}_{\beta}^{(i)} - {\sigma}_{\beta'}^{(i)} + 2u ) +\sum\limits_{a = 1}^{w_{i}} \left( {\varpi} ( {\sigma}_{\beta}^{(i)} - {\mu}_{a}^{(i)} - u ) +  {\varpi} ( - {\sigma}_{\beta}^{(i)} + {\mu}_{a}^{(i)} - u ) \right) \right)} + \\
  + \sum_{i=1}^{m-1}  \sum_{{\alpha} = 1}^{v_{i}} \sum_{{\beta}=1}^{v_{i+1}} \scriptstyle{\left( {\varpi} ( {\sigma}_{\alpha}^{(i)} - {\sigma}_{\beta}^{(i+1)} - u) + {\varpi} ( - {\sigma}_{\alpha}^{(i)} + {\sigma}_{\beta}^{(i+1)} - u) \right)}+  \\
 + \sum_{\beta = 1}^{v_{m}} \sum_{a = 1}^{w} \scriptstyle{\left(    {\varpi} ( {\sigma}_{\beta}^{(i)} - {\mu}_{a}^{(+)}) +  {\varpi} ( - {\sigma}_{\beta}^{(i)} + {\mu}_{a}^{(-)})  \right)} + \\
 + \sum_{i =m+1}^{m+n-1} \sum_{\beta = 1}^{v_{i}} \scriptstyle{\left(  \sum\limits_{{\beta}' = 1}^{v_{i}} {\varpi} ( {\sigma}_{\beta}^{(i)} - {\sigma}_{\beta'}^{(i)} - 2u ) +\sum\limits_{a = 1}^{w_{i}} \left( {\varpi} ( {\sigma}_{\beta}^{(i)} - {\mu}_{a}^{(i)} + u ) +  {\varpi} ( - {\sigma}_{\beta}^{(i)} + {\mu}_{a}^{(i)} + u ) \right) \right)} + \\
 + \sum_{i=m}^{m+n-1}  \sum_{{\alpha} = 1}^{v_{i}} \sum_{{\beta}=1}^{v_{i+1}} \scriptstyle{\left( {\varpi} ( {\sigma}_{\alpha}^{(i)} - {\sigma}_{\beta}^{(i+1)} + u) + {\varpi} ( - {\sigma}_{\alpha}^{(i)} + {\sigma}_{\beta}^{(i+1)} + u) \right)}  \\
 \label{eq:genyy2}
 \end{multline}
where
\beq
P_{i}(x) = \prod_{a=1}^{w_{i}} \, \left( x - {\mu}_{a}^{(i)} \right) \, , \qquad i = 1, \ldots , m-1, \pm , m+1, \ldots ,  n+m-1
\label{eq:pix}
\eeq
and ${\rm deg}P_{+} = {\rm deg}P_{-} = w$. 
\subsection{General Dynkin diagram}

The general Dynkin diagram of $\mathfrak{sl}(m|n)$ is characterized by a
collection of $p \geq 1$ integers $0 < l_{1} < l_{2} < \ldots < l_{p} < m+n$, 
labeling the chosen fermionic simple roots, where, 
for even $p = 2k$:
\beq
n = \sum_{i=1}^{p} (-1)^{i} l_{i} = d_{1} + d_{3} + \ldots + d_{2k-1} \ , 
\eeq
and for odd $p=  2k+1$:
\beq
m = \sum_{i=1}^{p} (-1)^{i-1} l_{i} = d_{1} + d_{3} + \ldots + d_{2k+1}  \ , 
\eeq
where $d_{0} = m+n- l_{p}$, $d_{i} = l_{p+1-i} - l_{p-i}$, $i = 1, \ldots, p-1$, $d_{p} = l_{1}$, so that all $d_{i} \geq 1$, and $\sum_i d_i = m+n$. 

In this paper we shall not discuss the Bethe/gauge correspondence for the general Dynkin diagrams of the $\mathfrak{gl}(m|n)$ superalgebra. We leave this as an exercise. 

\section{Supersymmetric gauge theory for superspin chain}

The first observation about \eqref{eq:genyy2} is that it is obtained by fusing two type $A$ quiver theories, $A_{m-1}$ and $A_{n-1}$, with the opposite values of the $u$-parameter. The fusing node $i=m$ is a $U(v_{m})$ ${\CalN}=(2,2)$ gauge theory which couples to both $A_{m-1}$ and $A_{n-1}$ theories. 

Here is the minimal construction, which we found in 2008 \footnote{Many thanks to E.~Ragoucy for helpful correspondence  and patient explanations of the results in \cite{Ragoucy:2007kg} then and ten years later} (the paper \cite{Orlando:2010uu} used the same construction in the $(m|n) = (2|1)$ case, albeit for $\mathfrak{sl}$ rather $\mathfrak{gl}$ superalgebra). 

Start with the $A_{m-1} \times A_{n-1}$ ${\CalN}=(4,4)$ theory with the gauge group $G_{l} \times G_{r}$ where $G_{l} = U(v_{1}) \times \ldots \times U(v_{m-1})$, 
$G_{r} = U(v_{m+1}) \times \ldots \times U(v_{m+n-1})$, the bi-fundamental hypermultiplets in $( {\bf v}_{i+1}, {\bar{\bf v}}_{i})$,  $i = 1, \ldots , m-2$, and $i = m+1, \ldots , m+n-2$,
and fundamental hypermultiplets $({\bar{\bf w}}_{i}, {\bf v}_{i})$, $i = 1, \ldots, m-1$, and  $i = m+1, \ldots, m+n-1$. Now let us turn on the twisted mass $u$ for the $U(1)_{\sf u}$ symmetry which acts as $U(1)_{u}$ on the fields of the $A_{m-1}$ sector and as $\overline{U(1)_{u}}$ on the fields of the $A_{n-1}$ sector (i.e. the opposite charges). As usual, we turn on the twisted masses for the maximal tori of the flavor symmetry
$U(w_{1}) \times \ldots \times U(w_{m-1}) \times U(w_{m+1}) \times \ldots \times U(w_{m+n-1})$. 

Now we couple this theory to the ${\CalN}=(2,2)$ gauge theory with the gauge group $U(v_{m})$, and the bi-fundamental chiral multiplets  $B_{m-1} \oplus {\tilde B}_{m-1}$ in 
$({\bar{\bf v}}_{m-1} , {\bf v}_{m}) \oplus ( {\bar{\bf v}}_{m}, {\bf v}_{m-1})$ and $B_{m} \oplus {\tilde B}_{m}$ in $({\bar{\bf v}}_{m+1} , {\bf v}_{m}) \oplus ( {\bar{\bf v}}_{m}, {\bf v}_{m+1})$ and the fundamental and anti-fundamental chirals $I_{m} \in ({\bar{\bf w}_{-}}, {\bf v}_{m})$ and $J_{m} \in ({\bar{\bf v}_{m}}, {\bf w}_{+})$, where
the vector spaces ${\bf w}_{\pm}$ have equal rank $w$. 

The matter fields couple to the ${\CalN}=(2,2)$ adjoint chirals at the $m-1$ and $m+1$ node through the superpotential (in addition to the superpotential inherited from the ${\CalN}=(4,4)$ theory):
\beq
{\delta}_{1}W = {\Tr}_{{\bf v}_{m}} \left(  B_{m-1} {\Phi}_{m-1} {\tilde B}_{m-1} \right)  - {\Tr}_{{\bf v}_{m}} \left(   B_{m} {\Phi}_{m+1} {\tilde B}_{m} \right)
\eeq
Thus, the chiral multiplets $B_{m-1}, {\tilde B}_{m-1}$ have the charge $-1$ under $U(1)_{\sf u}$ while $B_{m+1}, {\tilde B}_{m+1}$ have the charge $+1$ (recall that ${\Phi}_{i}$ has the charge $+2$ for $i < m$ and $-2$ for $i > m$).

\subsection{A family of theories}

The minimal choice above reproduces the equations \eqref{eq:genyy2}. 
However this choice lacks the rigidity one expects of the theory with the hidden $Y(\mathfrak{sl}(m|n))$ symmetry. Namely, the $U(1)_{\sf u}$ symmetry is a subgroup in $U(1)_{l} \times U(1)_{r}$, where $U(1)_{l,r}$
acts as $U(1)_{u}$ on the $A_{m-1}$ and on the $A_{n-1}$ portions, respectively, including the bifundamentals $(B_{m-1}, {\tilde B}_{m-1})$
and $(B_{m+1}, {\tilde B}_{m+1})$ (which are fundamental hypermultiplets from the point of view of $A_{m-1}$ and $A_{n-1}$ portions, respectively). 
One can therefore deform this theory by two twisted masses $u_{l}, u_{r}$, so that the theory we discussed so far would correspond to the
case $u_{l} + u_{r} = 0$. It is possible that such deformation also has an interesting Bethe/gauge dual (perhaps the generalized root systems of \cite{SSS:2018} would make an appearence, with ${\kappa}/(1-{\kappa}) = - u_{r}/u_{l}$). 

We propose another solidifier. Introduce the triplet $({\Phi}_{-}, {\Phi}_{0}, {\Phi}_{+})$ of $U(v_{m})$ adjoint  chiral multiplets, with the $U(1)_{\sf u}$ charges $+2, 0, -2$, respectively, and add the following terms to the superpotential:
\beq
{\delta}_{2}W = {\Tr}_{{\bf v}_{m}} \left( {\Phi}_{0} [{\Phi}_{+}, {\Phi}_{-}] - 
{\Phi}_{+}  B_{m-1}  {\tilde B}_{m-1} + {\Phi}_{-}  B_{m+1}  {\tilde B}_{m+1} \right)
\label{eq:d2w}
\eeq
and
\beq
{\delta}_{3} W = t_{1}\, {\Tr}_{{\bf v}_{m}} {\Phi}_{+} {\Phi}_{-}  +  t_{2} {\Tr}_{{\bf v}_{m}} {\Phi}_{0}^{2} \ .
\label{eq:lock}
\eeq
The $U(1)_{\sf u}$-symmetry allows one to add terms like $U({\Phi}_{0})$ with some gauge-invariant polynomial $U(x)$, or $\sum_{l} s_{l} {\Tr} \left( {\Phi}_{+} {\Phi}_{-}  {\Phi}_{0}^{l}\right)$, however our choices are limited by cubic polynomials as we wish to be able to lift these theories to renormalizable ${\CalN}=1$ theories in four dimensions (with the XXZ and XYZ-type Bethe duals). 

The term \eqref{eq:d2w} can be accompanied by the coupling ${\delta}_{4}W = {\Tr} {\Phi}_{0} IJ$ to yet another fundamental hypermultiplet $(I, J) \in ({\bar{\bf w}_{0}} , {\bf v}_{m}) \oplus ({\bar{\bf v}_{m}}, {\bf w}_{0})$. Neither $\Phi_0$ nor $(I,J)$ contribute to the effective twisted superpotential ${\tilde W}$ since ${\Phi}_{0}$ has charge $0$ under $U(1)_{\sf u}$ and $I$ and $J$ have the opposite charges (which can be absorbed into the twisted masses for $U(w_{0})$ flavor symmetry). 
The nice feature of the $({\Phi}_{0, \pm}, B_{m}, {\tilde B}_{m}, B_{m-1}, {\tilde B}_{m-1}, I,J)$ package is that its Higgs branch coincides with the moduli space of spiked instantons \cite{NekrasovF} which fit into a three dimensional variety (see \cite{Soibelman} for the recent work where
using these moduli spaces the representations of the cohomological Hall algebra are constructed). 
In the absence of the $(I,J)$-matter fields the corresponding Higgs branch is the moduli space of folded instantons \cite{Nbpscft} which we shall discuss in the next section. 

We should stress that only the ${\delta}_{3}W$ term provides the rigidity $u_{l}+u_{r}= 0$. Once $t_{1}= t_{2} = 0$ we can turn on both $u_{l}$ and $u_{r}$, leading to the equations describing the quantum cohomology, i.e. the spectrum of the twisted chiral ring: whenever $Q_{i}(x) = 0$, 
\beq
\begin{aligned}
& \quad  \frac{Q_{i-1}(x + u_{l})}{Q_{i-1}(x - u_{l})}\frac{Q_{i}(x-  2u_{l})}{Q_{i}(x + 2u_{l})}\frac{Q_{i+1}(x + u_{l})}{Q_{i+1}(x - u_{l})} = - {\qe}_{i}^{-1} \frac{P_{i}(x + u_{l})}{P_{i}(x - u_{l})} \, ,  \\
& \qquad\qquad\qquad\qquad \qquad\qquad\qquad\qquad \qquad\qquad\qquad\qquad \scriptstyle{1 \leq i \leq m-1} \, , \\
& 
\ \frac{Q_{m-1}(x + u_{l})}{Q_{m-1}(x - u_{l})}
\frac{Q_{m}(x-  2u_{l})}{Q_{m}(x + 2u_{l})}\frac{Q_{m+1}(x + u_{r})}{Q_{m+1}(x - u_{r})}\frac{Q_{m}(x-  2u_{r})}{Q_{m}(x + 2u_{r})} \frac{Q_{m}(x +  2u_{l} + 2u_{r})}{Q_{m}(x - 2u_{l}- 2u_{r})}= \\
& \qquad\qquad = - {\qe}_{m}^{-1} \frac{P_{+}(x)}{P_{-}(x)} \frac{P_{m}(x - u_{l}- u_{r})}{P_{m}(x +u_{l}+u_{r})} \, , \\
& \quad \, \frac{Q_{i-1}(x +u_{r})}{Q_{i-1}(x - u_{r})}\frac{Q_{i}(x -  2u_{r})}{Q_{i}(x + 2u_{r})}\frac{Q_{i+1}(x + u_{r})}{Q_{i+1}(x - u_{r})} = -  {\qe}_{i}^{-1} \frac{P_{i}(x + u_{r})}{P_{i}(x - u_{r})} \, , \\
& \qquad\qquad\qquad\qquad \qquad\qquad\qquad\qquad \qquad\qquad\qquad \qquad \scriptstyle{m < i \leq m+n-1} \, , \\
\label{eq:baesusy2}
\end{aligned}
\eeq
where ${\qe}_{i} = e^{2\pi \ii {\vartheta}_{i} - r_{i}}$'s are the Kahler moduli. 
The $t_{1} = t_{2} = 0$ locus  has a bonus feature in the form of a $U(1)_{R}$ symmetry, under which all the fundamentals except $(I,J)$ and bi-fundamentals have charge $0$, all the ${\Phi}_{i}$, ${\Phi}_{\pm}$ fields have charge $+1$, with ${\Phi}_{0}$ having charge $-1$, and $I, J$ having charge $+1$. This symmetry is preserved by the $\beta$-deformation:
\beq
{\Tr} {\Phi}_{0} [ {\Phi}_{+}, {\Phi}_{-}]  \longrightarrow 
e^{\beta} {\Tr} \left( {\Phi}_{0} {\Phi}_{+} {\Phi}_{-} \right) - e^{-\beta} {\Tr} \left( {\Phi}_{0} {\Phi}_{-} {\Phi}_{+} \right) \label{eq:betdef}
\eeq
Likewise, this $U(1)_{R}$ symmetry is restored in the limit where  both $t_{1}$ and $t_{2}$ go to infinity, i.e. ${\Phi}_{\pm}$ and ${\Phi}_{0}$ decouple.

The $U(1)_{R}$ symmetry can be used to define the topological field theory by $A$ twist. After the twist the fields ${\Phi}_{i}, {\Phi}_{\pm}, I, J$ become the $(1,0)$-forms on the worldsheet $\Sigma$, i.e. 
${\Phi}_{i} = {\Phi}_{i,z} dz \in {\Gamma} \left( {\rm End}({\CalV}_{i}) \otimes K_{\Sigma} \right)$, $I = I_{z}dz \in {\Gamma} \left( {\rm Hom}( {\bf w}_{0} , {\CalV}_{m}) \otimes K_{\Sigma} \right)$, while ${\Phi}_{0}$ becomes the section of ${\rm End}({\CalV}_{m}) \otimes {\CalT}_{\Sigma}$. The path integral localizes onto the solutions of the generalized Hitchin equations, which schematically read as follows:
\beq
{\nabla}_{\bar z} ( {\rm field} ) = \left( {\partial} W / {\partial} {\rm field} \right)^{\dagger}
\eeq
where by the field we mean the lowest component of the chiral multiplet after the twisting. 

When $\Sigma = D^{2}$ or ${\Sigma} = {\BC}$ one can further deform the theory by subjecting it to the two-dimensional $\Omega$-background. The path integral with the supersymmetric boundary conditions is expected to solve the quantum Knizhnik-Zamolodchikov equation based on superalgebras, cf. \cite{AOqm}. 

\section{Conclusions and future prospects}

Bethe/gauge correspondence between the finite-dimensional spin chains and two dimensional supersymmetric gauge theories (their anisotropic cousins corresponding to the three and four dimensional theories toroidally compactified to two dimensions) has a parallel correspondence between the quantum integrable systems with infinite-dimensional spaces of states, such as many-body systems, and the four (five, six) dimensional supersymmetric gauge theories subject to a two dimensional $\Omega$-background (times a circle or a torus) \cite{NS09,NPS}. The examples discussed in this paper are not an exception to that rule. Namely, there is a four-dimensional theory 
subject to a two dimensional $\Omega$-background, which corresponds to a many-body system based on superalgebra $\mathfrak{sl}(m|n)$. 
It was shown in \cite{Nbpscft} that the {\tt folded instanton} theory, i.e. a generalized gauge theory on the spacetime of the form: ${\BC} \times {\BC} \cup_{0} {\BC}$ (in other words, a union of the coordinate planes ${\BC}^{2}_{12}$ ($z_{3}= 0$) and ${\BC}^{2}_{23}$ ($z_{1}=0$) inside the three complex dimensional space ${\BC}^{3}$ with the coordinates $z_{1}, z_{2}, z_{3}$), with the local gauge groups $U(n)$ and $U(m)$ (and local matter content of the ${\CalN}=2^{*}$ theory), respectively, subject to the $\Omega$-deformations in ${\BC}^{1}_{1}$ and ${\BC}^{1}_{3}$ with the equivariant parameters
${\ve}_{1}$ and ${\ve}_{3}$, respectively, is a theory with the ${\CalN}=(2,2)$ super-Poincare invariance in two dimensions (i.e. in ${\BC}^{1}_{2}$). Its Bethe dual is the deformed elliptic Calogero-Moser system (the trigonometric version was studied in \cite{SV}): 
\begin{multline}
{\hat H} = - \frac{\kappa}{2} \sum_{i=1}^{n} \frac{{\partial}^{2}}{\partial x_{i}^2} - \frac{1-\kappa}{2} \sum_{j=1}^{m} \frac{{\partial}^{2}}{\partial y_{j}^2} + \\
+ \frac{\kappa}{1-\kappa} \sum_{i< i'}  {\wp} (x_{i} - x_{i'}) + \frac{1-\kappa}{\kappa} \sum_{j<j'} {\wp} (y_{j} - y_{j'}) + \sum_{i,j} {\wp} (x_{i} - y_{j}) \, , 
\label{eq:defcm}
\end{multline}
where
\beq
{\kappa} = \frac{{\ve}_{1}}{{\ve}_{1}+ {\ve}_{3}}
\eeq
It was shown in \cite{Nbpscft} that the partition function of the theory with the surface defect inserted at some point in ${\BC}^{1}_{2}$ (with the monodromy defect at $0 \in {\BC}^{1}_{1} \cup_{0} {\BC}^{1}_{3}$)
is the wavefunction of the quantum system \eqref{eq:defcm}. Specifically, such a partition function is obtained by integration over a ${\BZ}_{m+n}$-fixed locus in the moduli space of folded instantons, which is the space of solutions to the following system of equations:
\beq
\begin{aligned}
 & [ {\Phi}_{+}, {\Phi}_{-} ] = 0 \, , \\
 &  [ {\Phi}_{0}, {\Phi}_{+} ] +  B_{m+1}{\tilde B}_{m+1} = 0 \, , \\
 & [ {\Phi}_{0} , {\Phi}_{-}] + {\tilde B}_{m} B_{m} = 0 \\
 & {\Phi}_{-} B_{m+1} = {\Phi}_{+} {\tilde B}_{m} = {\tilde B}_{m+1} {\Phi}_{-} = B_{m} {\Phi}_{+} = 0 \ .
 \end{aligned}
 \eeq
We expect that a proper large $m,n$ limit of this model produces a super-version of the quantum intermediate long-wave equation, whose spectrum is determined from the  Bethe equations similar to \eqref{eq:baesusy2}. 

On the other hand, the surface defect of the folded instanton theory can be modelled on a two dimensional ${\CalN}=(2,2)$ gauged linear sigma model albeit on the worldsheet made out of two copies of ${\BC}^1$ (more specifically ${\BC}^{1}_{1}$ and ${\BC}^{1}_{3}$) glued at one point $0$. On either component the low-energy effective target space is the cotangent bundle to the complete flag variety, $T^{*}Fl(m,m-1,\ldots, 1)$ and $T^{*}Fl(n,n-1, \ldots, 1)$, respectively. In addition, there are degrees of freedom localized at $0$, which describe some interaction between the two sigma models. We expect the equivalence between the four dimensional and the two dimensional viewpoints on this system is a way to make contact with the discrete dynamics approach to Bethe ansatz of superalgebras of \cite{Kazakov:2007fy}. 

Finally, let us mention another extension of this work. In \cite{NPS} the $ADE$-type quiver gauge theories in four and five dimensions were analyzed using the $q$-character \cite{FR} observables, which were generalized to $qq$-characters in \cite{Nbpscft}. In \cite{Kimura:2017hez} the theories associated to the non-simply-laced algebras were constructed, together with the corresponding $qq$-characters. It must be possible to include the superalgebras into this picture as well, in particular to define the $qq$-characters for the Yangians and quantum affine algebras based on $\mathfrak{sl}(m|n)$. The surface defects in these theories will presumably carry the ${\CalN}=2$ structure in two and three dimensions that we described in this note.


\begin{thebibliography}{KLLSW}


\bibitem{Aganagic:2003xq} 
  M.~Aganagic, K.~A.~Intriligator, C.~Vafa and N.~P.~Warner,
  \emph{The Glueball superpotential},
  Adv.\ Theor.\ Math.\ Phys.\  {\bf 7}, no. 6, 1045 (2003)
  doi:10.4310/ATMP.2003.v7.n6.a4
  [hep-th/0304271].


\bibitem{Aganagic:2016jmx} 
  M.~Aganagic and A.~Okounkov,
 \emph{Elliptic stable envelope},
  arXiv:1604.00423 [math.AG].
  
\bibitem{AOqm}
  M.~Aganagic and A.~Okounkov,
 \emph{Quasimap counts and Bethe eigenfunctions},
  Moscow Math.\ J.\  {\bf 17}, no. 4, 565 (2017)
  [arXiv:1704.08746 [math-ph]].
      
  
  \bibitem{AlvarezGaume:1983ab} 
  L.~Alvarez-Gaume and D.~Z.~Freedman,
  \emph{Potentials for the Supersymmetric Nonlinear Sigma Model},
  Commun.\ Math.\ Phys.\  {\bf 91}, 87 (1983).
  doi:10.1007/BF01206053

  \bibitem{Arutyunov:2004vx} 
  G.~Arutyunov, S.~Frolov and M.~Staudacher,
  \emph{Bethe ansatz for quantum strings},
  JHEP {\bf 0410}, 016 (2004)
  doi:10.1088/1126-6708/2004/10/016
  [hep-th/0406256].


\bibitem{Beisert:2003yb} 
  N.~Beisert and M.~Staudacher,
  \emph{The ${\CalN}=4$ SYM integrable super spin chain},
  Nucl.\ Phys.\ B {\bf 670}, 439 (2003)
  doi:10.1016/j.nuclphysb.2003.08.015
  [hep-th/0307042].

  
  \bibitem{Beisert:2004hm} 
  N.~Beisert, V.~Dippel and M.~Staudacher,
  \emph{A Novel long range spin chain and planar ${\CalN}=4$ super Yang-Mills},
  JHEP {\bf 0407}, 075 (2004)
  doi:10.1088/1126-6708/2004/07/075
  [hep-th/0405001].
  
  \bibitem{Beisert:2006ez} 
  N.~Beisert, B.~Eden and M.~Staudacher,
 \emph{Transcendentality and Crossing},
  J.\ Stat.\ Mech.\  {\bf 0701}, P01021 (2007)
  doi:10.1088/1742-5468/2007/01/P01021
  [hep-th/0610251].
  
  
  \bibitem{Beisert:2006ib} 
  N.~Beisert, R.~Hernandez and E.~Lopez,
\emph{A Crossing-symmetric phase for $AdS_{5} x S^{5}$ strings},
  JHEP {\bf 0611}, 070 (2006)
  doi:10.1088/1126-6708/2006/11/070
  [hep-th/0609044].
  
  \bibitem{Beisert:2008tw} 
  N.~Beisert and P.~Koroteev,
  J.\ Phys.\ A {\bf 41}, 255204 (2008)
  doi:10.1088/1751-8113/41/25/255204
  [arXiv:0802.0777 [hep-th]].
  
  
  \bibitem{Belliard:2008di} 
  S.~Belliard and E.~Ragoucy,
\emph{Nested Bethe ansatz for 'all' closed spin chains},
  J.\ Phys.\ A {\bf 41}, 295202 (2008)
  doi:10.1088/1751-8113/41/29/295202
  [arXiv:0804.2822 [math-ph]].


  \bibitem{Dijkgraaf:2016lym} 
  R.~Dijkgraaf, B.~Heidenreich, P.~Jefferson and C.~Vafa,
  \emph{Negative Branes, Supergroups and the Signature of Spacetime},
  JHEP {\bf 1802}, 050 (2018)
  doi:10.1007/JHEP02(2018)050
  [arXiv:1603.05665 [hep-th]].
 
 
\bibitem{Dorey:2007xn} 
  N.~Dorey, D.~M.~Hofman and J.~M.~Maldacena,
  \emph{On the Singularities of the Magnon S-matrix},
  Phys.\ Rev.\ D {\bf 76}, 025011 (2007)
  doi:10.1103/PhysRevD.76.025011
  [hep-th/0703104 [HEP-TH]].

\bibitem{Drummond:2010km} 
  J.~M.~Drummond,
  \emph{Review of AdS/CFT Integrability, Chapter V.2: Dual Superconformal Symmetry},
  Lett.\ Math.\ Phys.\  {\bf 99}, 481 (2012)
  doi:10.1007/s11005-011-0519-4
  [arXiv:1012.4002 [hep-th]].
   
    
\bibitem{Frappat:1996pb} 
  L.~Frappat, P.~Sorba and A.~Sciarrino,
  \emph{Dictionary on Lie superalgebras},
  hep-th/9607161.
  

\bibitem{FR} E.~Frenkel, N.~Reshetikhin, \emph{The $q$-characters
 of representations of quantun affine algebras and deformations of 
 $W$-algebras}, arXiv:math/9810055v5 [math.QA]
  
 
  \bibitem{GS}
  A.~A.~Gerasimov and S.~L.~Shatashvili,
  \emph{Higgs Bundles, Gauge Theories and Quantum Groups},
  Commun.\ Math.\ Phys.\  {\bf 277}, 323 (2008)
  doi:10.1007/s00220-007-0369-1
  [hep-th/0609024].\\
    \emph{Two-dimensional gauge theories and quantum integrable systems},
  Proc.\ Symp.\ Pure Math.\  {\bf 78}, 239 (2008)
  [arXiv:0711.1472 [hep-th]].

   
  
\bibitem{Gorsky:1993pe} 
  A.~Gorsky and N.~Nekrasov,
 \emph{Hamiltonian systems of Calogero type and two-dimensional Yang-Mills theory},
  Nucl.\ Phys.\ B {\bf 414}, 213 (1994)
  doi:10.1016/0550-3213(94)90429-4
  [hep-th/9304047].
  
  

\bibitem{Gromov:2006dh} 
  N.~Gromov, V.~Kazakov, K.~Sakai and P.~Vieira,
  \emph{Strings as multi-particle states of quantum sigma-models},
  Nucl.\ Phys.\ B {\bf 764}, 15 (2007)
  doi:10.1016/j.nuclphysb.2006.11.018
  [hep-th/0603043].
  
  \bibitem{Gromov:2006cq} 
  N.~Gromov and V.~Kazakov,
  \emph{Asymptotic Bethe ansatz from string sigma model on ${\BS}^{3} \times {\BR}$},
  Nucl.\ Phys.\ B {\bf 780}, 143 (2007)
  doi:10.1016/j.nuclphysb.2007.03.025
  [hep-th/0605026].

  
  
  \bibitem{Gromov:2007fn} 
  N.~Gromov, V.~Kazakov and P.~Vieira,
  \emph{Classical limit of Quantum Sigma-Models from Bethe Ansatz},
  PoS SOLVAY {\bf }, 005 (2006)
  doi:10.22323/1.038.0005
  [hep-th/0703137 [HEP-TH]].
  
  
  
  
\bibitem{Hutsalyuk:2017tcx} 
  A.~Hutsalyuk, A.~Liashyk, S.~Z.~Pakuliak, E.~Ragoucy and N.~A.~Slavnov,
 \emph{Scalar products of Bethe vectors in the models with $\mathfrak{gl}(m|n)$ symmetry},
  Nucl.\ Phys.\ B {\bf 923}, 277 (2017).
  doi:10.1016/j.nuclphysb.2017.07.020


  
  \bibitem{Janik:2006dc} 
  R.A.Janik,
 \emph{The $AdS_{5} \times S^{5}$ superstring worldsheet $S$-matrix and crossing symmetry},
  Phys.\ Rev.\ D {\bf 73}, 086006 (2006)
  doi:10.1103/PhysRevD.73.086006
  [hep-th/0603038].



\bibitem{Kazakov:2004qf} 
  V.A.Kazakov, A.Marshakov, J.A.Minahan and K.Zarembo,
  \emph{Classical/quantum integrability in AdS/CFT},
  JHEP {\bf 0405}, 024 (2004)
  doi:10.1088/1126-6708/2004/05/024
  [hep-th/0402207].

 
  \bibitem{Kazakov:2007fy} 
  V.~Kazakov, A.~S.~Sorin and A.~Zabrodin,
 \emph{Supersymmetric Bethe ansatz and Baxter equations from discrete Hirota dynamics},
  Nucl.\ Phys.\ B {\bf 790}, 345 (2008)
  doi:10.1016/j.nuclphysb.2007.06.025
  [hep-th/0703147 [HEP-TH]].
  
 

    \bibitem{Kimura:2017hez} 
  T.~Kimura and V.~Pestun,
  \emph{Fractional quiver $W$-algebras},
  Lett.\ Math.\ Phys.\  {\bf 108}, no. 11, 2425 (2018)
  doi:10.1007/s11005-018-1087-7
  [arXiv:1705.04410 [hep-th]].
    
  
   \bibitem{KR} A.N.~Kirillov, N.~Reshetikhin, \emph{Representations of Yangians and multiplicities of
the inclusion of the irreducible components of the tensor product of representations
of simple Lie algebras} , (Russian) Zap. Nauchn. Sem. LOMI {\bf 160} (1987), Anal. Teor. Chisel i Teor. Funktsii. 8, 211Ð221, 301;
translation in J. Soviet Math. {\bf 52} (1990), no. 3, 3156Ð3164
  
  \bibitem{Kulish:1986} P.~P.~Kulish, J.~Sov.~Math {\bf 35} (1986) 2648 \ .
  %
  
  \bibitem{MO} D.~Maulik and A.~Okounkov, \emph{Quantum groups and quantum cohomology}, arXiv:1211.1287 



\bibitem{Migdal}
A.~A.~Migdal, Zh. Eksp. Teor. Fiz. 69, 810 (1975)
  
  
\bibitem{Minahan:2002ve} 
  J.~A.~Minahan and K.~Zarembo,
\emph{The Bethe ansatz for ${\CalN}=4$ superYang-Mills},
  JHEP {\bf 0303}, 013 (2003)
  doi:10.1088/1126-6708/2003/03/013
  [hep-th/0212208].
  
  
 
      
  \bibitem{Mikhaylov:2015nsa} 
  V.~Mikhaylov,
\emph{Analytic Torsion, 3d Mirror Symmetry And Supergroup Chern-Simons Theories},
  arXiv:1505.03130 [hep-th].
  
 \bibitem{Mikhaylov:2014aoa} 
  V.~Mikhaylov and E.~Witten,
  \emph{Branes And Supergroups},
  Commun.\ Math.\ Phys.\  {\bf 340}, no. 2, 699 (2015)
  doi:10.1007/s00220-015-2449-y
  [arXiv:1410.1175 [hep-th]].
  
  
  \bibitem{MNS} 
  G.~W.~Moore, N.~Nekrasov and S.~Shatashvili,
  \emph{Integrating over Higgs branches},
  Commun.\ Math.\ Phys.\  {\bf 209}, 97 (2000)
  doi:10.1007/PL00005525
  [hep-th/9712241].

  \bibitem{Nakajima} H.~Nakajima, \emph{Quiver varieties and finite dimensional representations of quantum affine algebras
}, arXiv:math/9912158 


  \bibitem{Nekrasov:2009uh} 
  N.~A.~Nekrasov and S.~L.~Shatashvili,
  \emph{Supersymmetric vacua and Bethe ansatz},
  Nucl.\ Phys.\ Proc.\ Suppl.\  {\bf 192-193}, 91 (2009)
  doi:10.1016/j.nuclphysbps.2009.07.047
  [arXiv:0901.4744 [hep-th]].
  
  
  \bibitem{NS09} 
  N.~A.~Nekrasov and S.~L.~Shatashvili,
  \emph{Quantization of Integrable Systems and Four Dimensional Gauge Theories},
  doi:10.1142/9789814304634$\_$0015
  arXiv:0908.4052 [hep-th].
\bibitem{NPS} 
  N.~Nekrasov, V.~Pestun and S.~Shatashvili,
  \emph{Quantum geometry and quiver gauge theories},
  Commun.\ Math.\ Phys.\  {\bf 357}, no. 2, 519 (2018)
  doi:10.1007/s00220-017-3071-y
  [arXiv:1312.6689 [hep-th]].


  \bibitem{NS} N.~Nekrasov, S.~Sethi, in progress
  
  
\bibitem{Nbpscft} N.~Nekrasov, \emph{BPS/CFT correspondence:
    non-perturbative Dyson-Schwinger equations and qq-characters},
  JHEP {\bf 1603}, 181 (2016) [arxiv:1512.05388 [hep-th]\\
  $\underline{~~~~~~~~~~}$, \emph{BPS/CFT correspondence II:
    Instantons at crossroads, Moduli and Compactness Theorem},
  arXiv:1608.07272 [hep-th]\\
  $\underline{~~~~~~~~~~}$, \emph{BPS/CFT correspondence III:
    Gauge Origami Partition Function and $qq$-characters},
  arXiv:1701.00189 [hep-th]\\
  $\underline{~~~~~~~~~~}$, \emph{BPS/CFT correspondence IV: sigma models and defects in gauge theory},
  doi:10.1007/s11005-018-1115-7
  arXiv:1711.11011 [hep-th]\\ 
  $\underline{~~~~~~~~~~}$, \emph{BPS/CFT correspondence V: BPZ and KZ equations from $qq$-characters},
  arXiv:1711.11582 [hep-th]\\
  N.~Nekrasov and N.~S.~Prabhakar,   \emph{Spiked Instantons from Intersecting D-branes},
  Nucl.\ Phys.\ B {\bf 914}, 257 (2017)
  doi:10.1016/j.nuclphysb.2016.11.014
  [arXiv:1611.03478 [hep-th]].
  
\bibitem{Okounkov:2015spn} 
  A.~Okounkov,
  \emph{Lectures on K-theoretic computations in enumerative geometry},
  arXiv:1512.07363 [math.AG].
  
  
 \bibitem{Orlando:2010uu} 
  D.~Orlando and S.~Reffert,
  \emph{Relating Gauge Theories via Gauge/Bethe Correspondence},
  JHEP {\bf 1010}, 071 (2010)
  doi:10.1007/JHEP10(2010)071
  [arXiv:1005.4445 [hep-th]].
 
  

\bibitem{Polyakov:2005ss} 
  A.~M.~Polyakov,
  \emph{Supermagnets and sigma models},
  doi:10.1142/9789812773784$\_$0036, 
  hep-th/0512310.
  

  \bibitem{Ragoucy:2007kg} 
  E.~Ragoucy and G.~Satta,
 \emph{Analytical Bethe Ansatz for closed and open ${\mathfrak{gl}}(M|N)$ super-spin chains in arbitrary representations and for any Dynkin diagrams},
  JHEP {\bf 0709}, 001 (2007)
  doi:10.1088/1126-6708/2007/09/001
  [arXiv:0706.3327 [hep-th]].
  
  

\bibitem{Soibelman} 
  M.~Rapcak, Y.~Soibelman, Y.~Yang and G.~Zhao,
  \emph{Cohomological Hall algebras, vertex algebras and instantons},
  arXiv:1810.10402 [math.QA].


  
\bibitem{Reshet} N.~Reshetikhin, \emph{The spectrum of the transfer matrices connected with Kac-Moody algebras},
Lett. Math. Phys. 14 (10): 235-246, 1987   \ . 
  
  
  
 \bibitem{RW} N.~Yu.~Reshetikhin and P.~B.~Wiegmann, Phys. Lett. {\bf B}189 (1987) 125  .


  
\bibitem{Sakai:2007rk} 
  K.~Sakai and Y.~Satoh,
  \emph{Origin of dressing phase in ${\CalN}=4$ super Yang-Mills},
  Phys.\ Lett.\ B {\bf 661}, 216 (2008)
  doi:10.1016/j.physletb.2008.02.015
  [hep-th/0703177].
  
  \bibitem{SSS:2018}
  S.~Sahi, H.~Salmasian and V.~Serganova, \emph{Capelli eigenvalue problem for Lie superalgebras and supersymetric polynominals}
 [arXiv:1807.07340 [math.RT]].
 
 
 
\bibitem{Schultz} C.~L.~Schultz, Physica {\bf A}122 (1983) 71 \ .  



 \bibitem{Serganova:2018}
 V.~Serganova, lecture at the meeting \emph{Representation Theory, Mathematical Physics and Integrable Systems}, on June 7,  2018 at the Centre International de Rencontres MathŽmatiques (Marseille, France),
 {\tt https://www.youtube.com/watch?v=vPNC4A6MimA}

  
  \bibitem{SV}
  A.~Sergeev and A.~Veselov, \emph{Deformed quantum Calogero-Moser problems and Lie superalgebras},  Comm.
Math. Phys., 245(2):249Ð278, 2004 \\
A.~Sergeev and A.~Veselov, \emph{Generalised discriminants, deformed Calogero-Moser-Sutherland operators and
super-Jack polynomials}, Adv. Math., 192(2):341Ð375, 2005.
 
 
  
  \bibitem{Vafa:2001qf} 
  C.~Vafa,
  \emph{Brane / anti-brane systems and $U(N|M)$ supergroup},
  hep-th/0101218.
  
  \bibitem{Varagnolo}
  M.~Varagnolo, \emph{Quiver Varieties and Yangians}, arXiv:math/0005277
  
  \bibitem{Witten:2d}
  E.~Witten, Comm. Math. Phys. 141 (1991) 153, https://doi.org/10.1007/BF02100009 
  
  \bibitem{Witten:1992xu} 
  E.~Witten,
 \emph{Two-dimensional gauge theories revisited},
  J.\ Geom.\ Phys.\  {\bf 9}, 303 (1992)
  doi:10.1016/0393-0440(92)90034-X
  [hep-th/9204083].

  
    
  \end{thebibliography}
\end{document}